\documentclass[a4paper,fleqn,usenatbib]{mnras}

\usepackage[T1]{fontenc}
\usepackage{ae,aecompl}

\usepackage{graphicx}	% Including figure files
\usepackage{amsmath}	% Advanced maths commands
\usepackage{amssymb}	% Extra maths symbols
\usepackage{color}

\title[Structural analysis of the Sextans dwarf spheroidal galaxy]{Structural analysis of the Sextans dwarf spheroidal galaxy}

\author[T. A. Roderick et al.]{
T. A. Roderick,$^{1}$\thanks{E-mail: Tammy.Roderick@anu.edu.au}
H. Jerjen,$^{1}$
G. S. Da Costa,$^{1}$
A. D. Mackey$^{1}$
\\
$^{1}$Research School of Astronomy and Astrophysics, Australian National University, Canberra, ACT 2611, Australia\\
}

\date{Accepted XXX. Received YYY; in original form ZZZ}

\pubyear{2015}

\begin{document}
\label{firstpage}
\pagerange{\pageref{firstpage}--\pageref{lastpage}}
\maketitle

% Abstract of the paper
\begin{abstract}
We present wide-field $g$ and $i$ band stellar photometry of the Sextans dwarf spheroidal galaxy and its surrounding area out to four times its half-light radius ($r_h=695\,$pc), based on images obtained with the Dark Energy Camera at the 4-m Blanco telescope at CTIO.  We find clear evidence of stellar substructure associated with the galaxy, extending to a distance of $82\arcmin$ (2\,kpc) from its centre. We perform a statistical analysis of the over-densities and find three distinct features, as well as an extended halo-like structure, to be significant at the $99.7\%$ confidence level or higher. Unlike the extremely elongated and extended substructures surrounding the Hercules dwarf spheroidal galaxy, the over-densities seen around Sextans are distributed evenly about its centre, and do not appear to form noticeable tidal tails.  Fitting a King model to the radial distribution of Sextans stars yields a tidal radius $r_t =83.2\arcmin\pm7.1\arcmin$ (2.08$\pm$0.18\,kpc), which implies the majority of detected substructure is gravitationally bound to the galaxy.  This finding suggests that Sextans is not undergoing significant tidal disruption from the Milky Way, supporting the scenario in which the orbit of Sextans has a low eccentricity.
\end{abstract}

% Select between one and six entries from the list of approved keywords.
% Don't make up new ones.

%%%%%%%%%%%%%%%%%%%%%%%%%%%%%%%%%%%%%%%%%%%%%%%%%%

%%%%%%%%%%%%%%%%% BODY OF PAPER %%%%%%%%%%%%%%%%%%

\section{Introduction}
\label{sec:intro}
The dawn of the survey telescope era has led to a rapid increase in discoveries of Milky Way (MW) satellite galaxies \citep[most recently:][]{2014ApJ...786L...3L,2015ApJ...805..130K,2015ApJ...799...73K,2015ApJ...807...50B,2015ApJ...804L..44K,2015arXiv150707564L,2015arXiv150803622T}, spurring a flurry of research into low mass galaxy formation and testing of the galaxy-scale predictions of $\Lambda$CDM cosmological models.  There has been much discussion on the origin of these satellites, leading to questions of whether or not they are primordial in nature, or the tidal remnants of an interaction between the MW and another galaxy \citep{1999ApJ...524L..19M,Kroupa:2005iy,2005Natur.433..389D,2011ASL.....4..297D,2014MNRAS.442.2362P,2014MNRAS.442.2419Y,Pawlowski:2015ub}.

Kinematic studies of the MW satellites have revealed them to have a high dark matter content (most particularly the ultra-faint systems) \citep[e.g.][]{2007ApJ...670..313S,2007MNRAS.380..281M,2011ApJ...733...46S,2015arXiv150407916K,2015ApJ...810...56K}. The velocity dispersion measured in these pressure-dominated systems suggests they are more massive than the visible baryonic matter can account for.  However, this assumes the systems are in dynamic equilibrium.  Where a system is interacting with its host, it can undergo significant tidal stirring \citep{2012ApJ...751...61L} resulting in kinematic samples being contaminated by unbound stars \citep{2007MNRAS.378..353K}.  Many of the MW satellites possess substructure indicative of tidal stirring, and in some cases extreme tidal stripping, for example: Sagittarius \citep{1994Natur.370..194I,2013AJ....145..163N}, Ursa Minor \citep{2003AJ....125.1352P}, Ursa Major II \citep{2007ApJ...670..313S}, Carina \citep{Battaglia:2012it,2014MNRAS.444.3139M},  and Fornax \citep{2004AJ....127..832C}.  If the majority of MW satellite galaxies are undergoing some form of tidal interaction, their kinematic samples may well be affected and it will be necessary to review the estimates of their dark matter content.

Recently, the Hercules dwarf spheroidal has been demonstrated to possess extended stellar substructure, with over-densities found out to 5.8 times the half-light radius from the galaxy's centre \citep{Roderick:2015jj}.  Previous studies of this satellite have suggested that tidal disruption is a likely scenario \citep{2007ApJ...668L..43C,2009ApJ...704..898S,2012MNRAS.425L.101D}, however the full extent of the stellar debris around the galaxy had not been realised.  In addition to the discovery of these distant stellar over-densities, a strong correlation was noted between the position of the over-densities, the position of blue horizontal branch (BHB) stars, and the orbital path suggested by \citet{2010ApJ...721.1333M}. This study illustrates the usefulness of new generation wide-field imaging to scrutinise the outskirts of known MW satellite galaxies, in order to better understand the extent and nature of their stellar structure.

Following this line of reasoning, we present results of a large scale photometric study of the Sextans dwarf spheroidal galaxy. Sextans (see Table \ref{tab:sextans}), discovered by \citet{1990MNRAS.244P..16I}, was the eighth MW dwarf satellite galaxy to be found.  Like a typical MW dwarf, Sextans is old \citep[$\>12$ Gyr,][]{1991AJ....101..892M} and metal poor ([Fe/H]$\simeq-1.9$, \cite{2011ApJ...727...78K}), with a high mass-to-light ratio \citep[$M/L\approx 97$,][]{2009MNRAS.394L.102L}). Tidal stirring has been suggested as a possible cause of the large observed velocity dispersion \citep{1994MNRAS.269..957H}.  There has also been discussion on whether or not there are substructures at the centre of Sextans, with the observation of kinematically cold structures in the galaxy's centre \citep{2004MNRAS.354L..66K,2006ApJ...642L..41W,2011MNRAS.411.1013B}, and the suggestion of a dissolved star cluster at its core \citep{2012ApJ...759..111K}.  The Sextans dwarf also contains a substantial population of blue straggler stars (BSS) \citep[or possibly intermediate age main sequence stars formed 2-6 Gyr ago,][]{2003AJ....126.2840L}, with the brighter stars being more centrally concentrated than the fainter part of the population \citep{2003AJ....126.2840L}.  An extensive study performed by \citet{2009ApJ...703..692L} revealed that, although the majority of star formation in Sextans occurred $>11$ Gyr ago, there are slight differences in the star formation history between the different regions identified in Sextans (especially between inner and outer populations).  Their conclusion was that the primary driver for the radial stellar population gradient seen in this galaxy is the star formation history.  

Given the interesting kinematic features and star formation history of Sextans, we have performed a wide-field photometric study in order to gain a better understanding of what is happening in the outskirts of this galaxy. Our data set consists of five fields taken with the Dark Energy Camera (DECam) on the 4m Blanco telescope, at Cerro Tololo in Chile.  We perform our analysis with the intention of investigating the nature of Sextans stellar structure, and ascertaining the extent to which this system is being tidally perturbed by the MW potential.  Section \ref{sec:obs} details our observations as well as the data reduction and photometric process.  In Section \ref{sec:fgd} we illustrate our method for discriminating between Sextans stars and the MW halo stars, and we describe our method for creating a spatial map in Section \ref{sec:detect}.  In Section \ref{sec:profile} we perform an analysis of the structural parameters of Sextans. Our main analysis of the structure of Sextans is detailed in Section \ref{sec:identify}, using techniques analogous to those of \citet{Roderick:2015jj}.  This section also includes a brief analysis of the centre of Sextans.  Finally in Section \ref{sec:disc} we discuss our results and summarise our findings.

\begin{table}
\begin{center}
\caption{Fundamental Parameters of Sextans}
\begin{tabular}{llc}
\hline\hline
Parameter & Value & Ref.$^a$\\
\hline
RA (J2000) & 10:13:03.0 & 1 \\
DEC (J2000) & $-$1:36:53 & 1 \\
$l$ & $243.5^\circ$ & 1\\
$b$ & $42.3^\circ$ & 1\\
$D_\odot$ & $86\pm4$\,kpc & 2 \\
$(m-M)_0$ & $19.67 \pm 0.1$ & 2 \\
$\sigma$ & $7.9 \pm 1.3$ km s$^{-1}$ & 3\\
$v_\odot$ & $224.3 \pm 0.1$ km s$^{-1}$ & 3\\
$r_h$ & 695 pc & 4 \\
$M_V$ & $-9.3\pm0.5$\,mag & 4\\
{\rm [Fe/H]}& $-1.93 \pm 0.01$\,dex & 5\\
$\theta$ & $56.7^\circ \pm 2.8^\circ$ & 6\\
$\epsilon$ & $0.29 \pm 0.03$ & 6\\
\hline\hline
\label{tab:sextans}
\end{tabular}
\end{center}
\textbf{$^a$References:} (1) \citet{1990MNRAS.244P..16I}, (2) \citet{2009ApJ...703..692L}, (3) \citet{2009AJ....137.3109W}, (4) \citet{1995MNRAS.277.1354I}, (5) \citet{2011ApJ...727...78K}, (6) This work
\end{table}

\section{Observations and Data Reduction}
\label{sec:obs}
Observations were conducted on 2013 February 15 using DECam at the CTIO 4m Blanco telescope, as part of observing proposal 2013A-0617 (PI: D. Mackey).  The total DECam field-of-view is 3 square degrees, formed by a hexagonal mosaic of 62 2K$\times$4K CCDs, each with a pixel scale of $0\farcs27$/pix.

\begin{figure}
\centering
\includegraphics[width=\columnwidth]{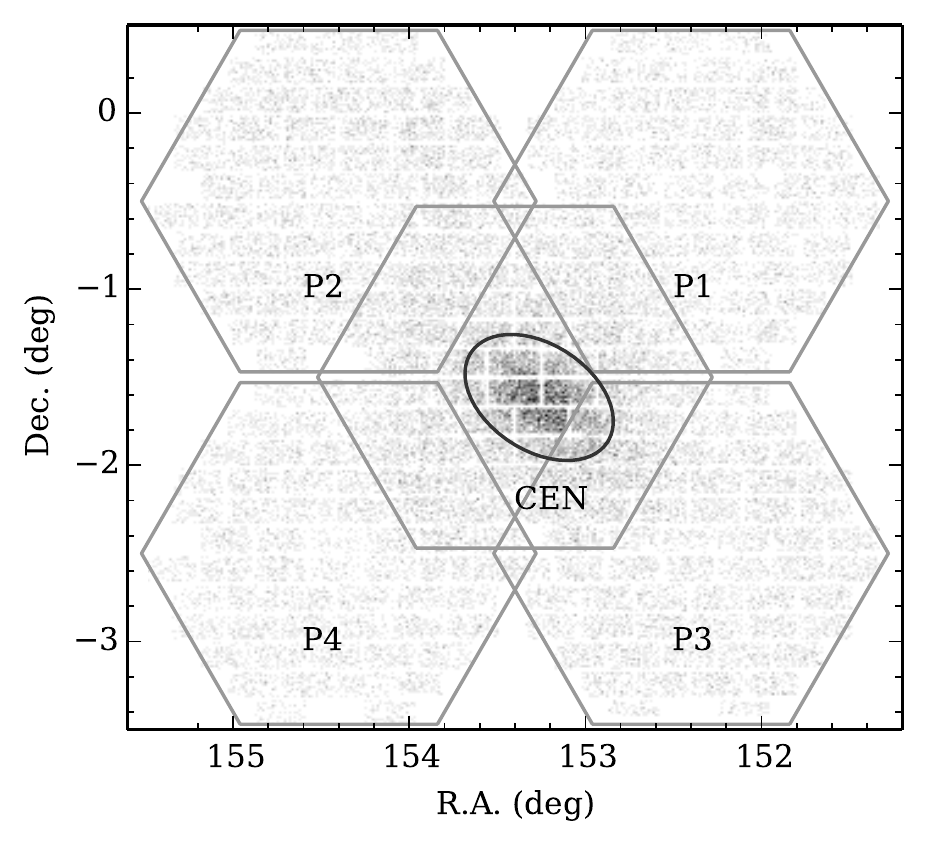}
\caption{Stellar distribution in five separate DECam pointings around the Sextans dwarf spheroidal galaxy.  The grey ellipse outlines the orientation and half-light radius of the galaxy.  Pointings are labelled CEN to P4 (outlined in light grey), and correspond to Table \ref{tab:data}. The slight tile pattern visible here is a result of the inter-chip gaps in the CCD mosaic.  Several CCDs in the mosaic were not functioning properly at the time of observation and have been masked out of the data set.  These are visible as irregular white gaps in the hexagonal pattern.}
\label{fig:data}
\end{figure}

The data set consists of five separate DECam pointings: one centred on Sextans itself (CEN) and four arranged symmetrically about the centre (P1-P4). The central coordinates for each of the outer fields are offset from the central coordinates of the centre field by $1^\circ$ in each R.A. and Dec. (see Figure \ref{fig:data}).  This configuration encompasses four times the half-light radius of Sextans \citep[27.6\arcmin,][]{1995MNRAS.277.1354I}, and provides over-lapping regions between each of the outer fields and the centre for photometric calibration. Each field was imaged $3 \times 300$s in $g$-band and $i$-band filters, providing total integration times of 900s in each filter for each field.  Table \ref{tab:data} details the central coordinates and average seeing for each field.

The images were processed using the DECam community pipeline\footnote{http://www.ctio.noao.edu/noao/content/Dark-Energy-Camera-DECam}  \citep{2014ASPC..485..379V}, which included sky-background subtraction, WCS solution fitting and the co-addition of images into a single deep frame. The final product was a multi-extension FITS file for each filter of each field, containing the processed and co-added image-stacks sliced into nine separate extensions or tiles.

\begin{table*}
\begin{center}
\caption{Properties of observations of Sextans taken as part of observing proposal 2013A-0617.}
\label{tab:data}
\begin{tabular}{lccccccc}
\hline\hline
Field & Coordinates & Filter & Average Seeing & \multicolumn{2}{c}{Completeness} & Zero Point & Colour Term \\
  &  (J2000) & & ($\arcsec$) & (90\% ) & (50\%) & ($ZP$) & ($c$) \\
\hline
CEN & 10:13:03.0     -01:31:53&$g$ & 1.07 & 24.75&24.97 &31.177&0.060\\
 & &$i$ &0.97 &24.01& 24.26& 31.109&0.070\\
P1 &10:09:03.0     -00:31:53 &$g$  &1.16 & 24.65& 24.85 &31.248&0.055\\
 & &$i$ &1.10 & 23.82&  24.11 &31.161&0.072\\
P2 &10:17:03.0     -00:31:53 &$g$  &1.25 & 24.60&  24.78 &31.145&0.043\\
 & &$i$ & 1.11  & 23.81& 24.09 &31.450&0.061\\
P3 &10:09:02.8     -02:31:53  &$g$  & 1.33& 24.52& 24.73  &31.243&0.054\\
 & &$i$ &1.38 & 23.50& 23.78 &31.003&0.070\\
P4 & 10:17:03.3     -02:31:53&$g$  &1.43 & 24.39& 24.59 &31.136&0.061\\
 & &$i$ & 1.33& 23.50& 23.80 &31.246&0.071\\
 \hline \hline
\end{tabular}
\end{center}
\end{table*}

\subsection{Photometry and Star/Galaxy Separation}
Photometry and star/galaxy separation were conducted with the same pipeline used by \citet{Roderick:2015jj}.  WeightWatcher\footnote{http://www.astromatic.net/software/weightwatcher} \citep{2008ASPC..394..619M} was used in conjunction with the weight maps produced by the community pipeline to mask out non-science pixels in preparation for aperture photometry to be performed using Source Extractor\footnote{http://www.astromatic.net/software/sextractor} \citep{1996A&AS..117..393B}.

Source Extractor was run in two passes on each image. An initial shallow pass was performed to estimate the average FWHM ($\overline{F}$) of bright point-like sources in the field, where a point-like source was defined as being circular and away from other sources and CCD edges.  This information then provided the input for a second deeper pass, where aperture photometry was carried out within $1\times \overline{F}$ and  $2\times \overline{F}$.  Performing the photometry separately on each of the nine tiles across all five fields allowed the apertures to vary with changes in the point-spread-function across the large field-of-view. 

\begin{figure}
\centering
\includegraphics[width=\columnwidth]{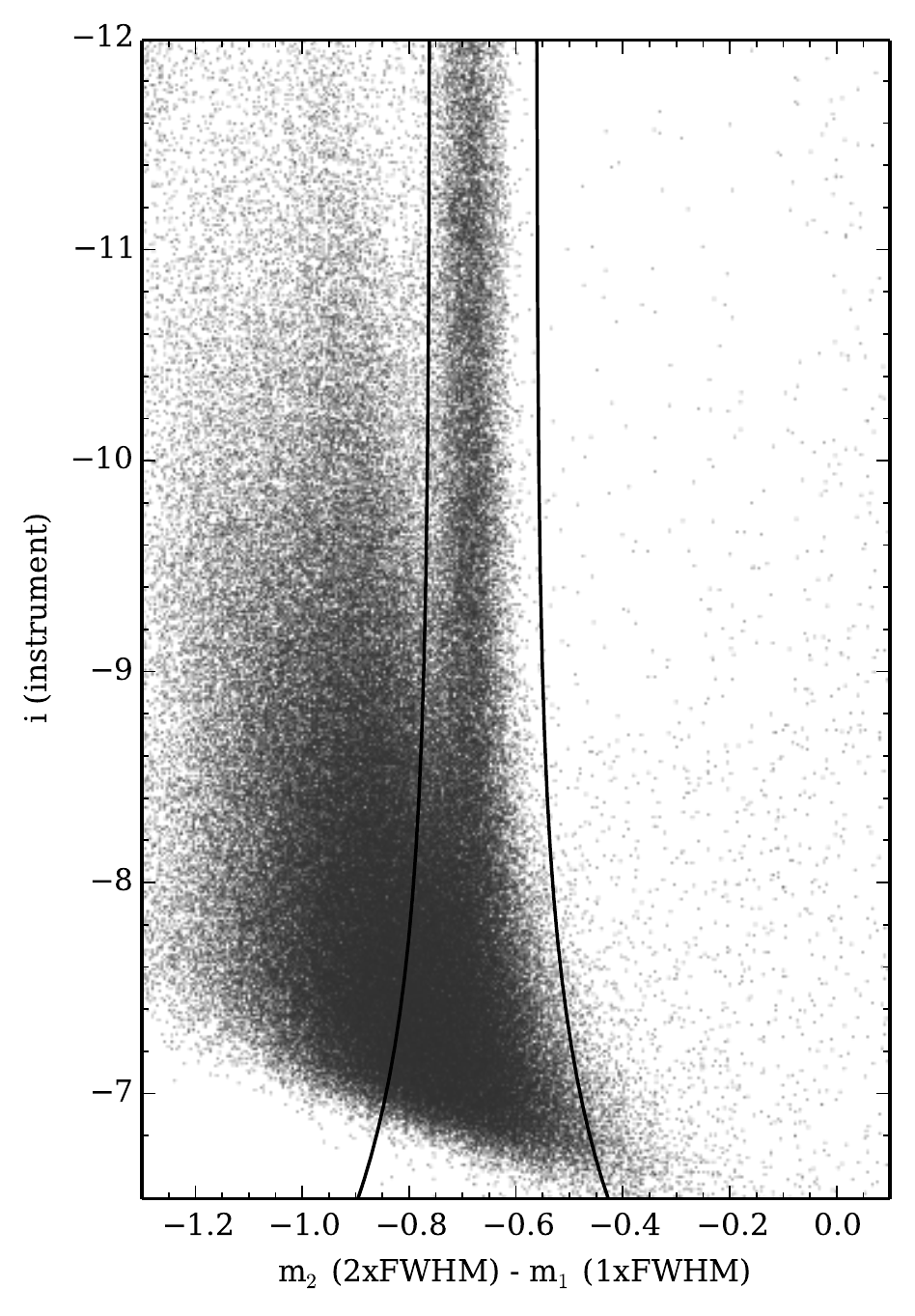}
\caption{Graphical interpretation of our star/galaxy separation process using the difference of magnitude between apertures for objects detected in P1. This figure shows the stellar locus, as well as a flare or plume of galaxies to the left and other spurious detections. The solid black lines denote the region inside which objects were classified as stars. They are catalogued and coordinate matched to their corresponding $g$-band counterpart.}
\label{fig:stargal}
\end{figure}

Star/galaxy separation was performed on the $i$-band image catalogues in the same manner as \citet{Roderick:2015jj}. Due to the variation in seeing between fields, the Source Extractor star/galaxy flag proved unreliable; some objects in fields of poorer seeing were unambiguously classified as stars, while in the overlapping region of the central field with best seeing, the same objects were unambiguously classified as galaxies. The aperture magnitude-difference method described by \citet{Roderick:2015jj} was adopted instead, since investigation of the overlapping regions between fields showed that it was able to consistently classify stellar objects. As demonstrated in Figure \ref{fig:stargal}, stars show a consistent flux ratio when measured by different sized apertures, enabling them to be separated from non-stellar objects. For each field, the stellar locus was modelled with an exponential which was reflected and shifted to encompass and select objects in the locus. Note that the shape of the exponential allows for increasing photometric uncertainties at faint magnitudes. The resulting stellar catalogue for each field was then cross-matched according to the WCS sky coordinates with the corresponding $g$-band catalogue, using the Stilts command line package \citep{2011ascl.soft05001T}. This produced one catalogue of stellar sources for each of the five fields. As a further step to ensure the consistency of star/galaxy classification between fields, this process was performed iteratively by checking the overlapping regions for consistent numbers of stars. The 90\% photometric completeness level (discussed later) of the poorest quality field was adopted as the magnitude limit for the entire stellar catalogue, to ensure even depth across fields, and a comparison made between the overlapping regions of each outer field with the central field. The selection region encompassing the stellar locus in Figure \ref{fig:stargal} was adjusted until the star counts in the overlapping regions were consistent to within poisson noise. This ensured a smooth transition between fields with consistent star/galaxy separation optimised for the quality of each image.

\subsection{Photometric Calibration and Completeness}

Once each catalogue was complete, it was individually calibrated to the SDSS photometric system.  This was achieved by matching objects from the catalogue to stars from SDSS data release 10 \citep[SDSS DR10:][]{2014ApJS..211...17A}.  By comparing the SDSS magnitude of each star to its instrumental magnitude within the $2\times \overline{F}$ aperture, photometric zero points ($ZP$) and colour terms ($c$) were determined using a least-squares fit; colour defined as $(g-i)_{inst}$.  The zero points and colour terms for each catalogue are summarised in Table \ref{tab:data}.

\begin{figure*}
\centering
\includegraphics[width=\textwidth]{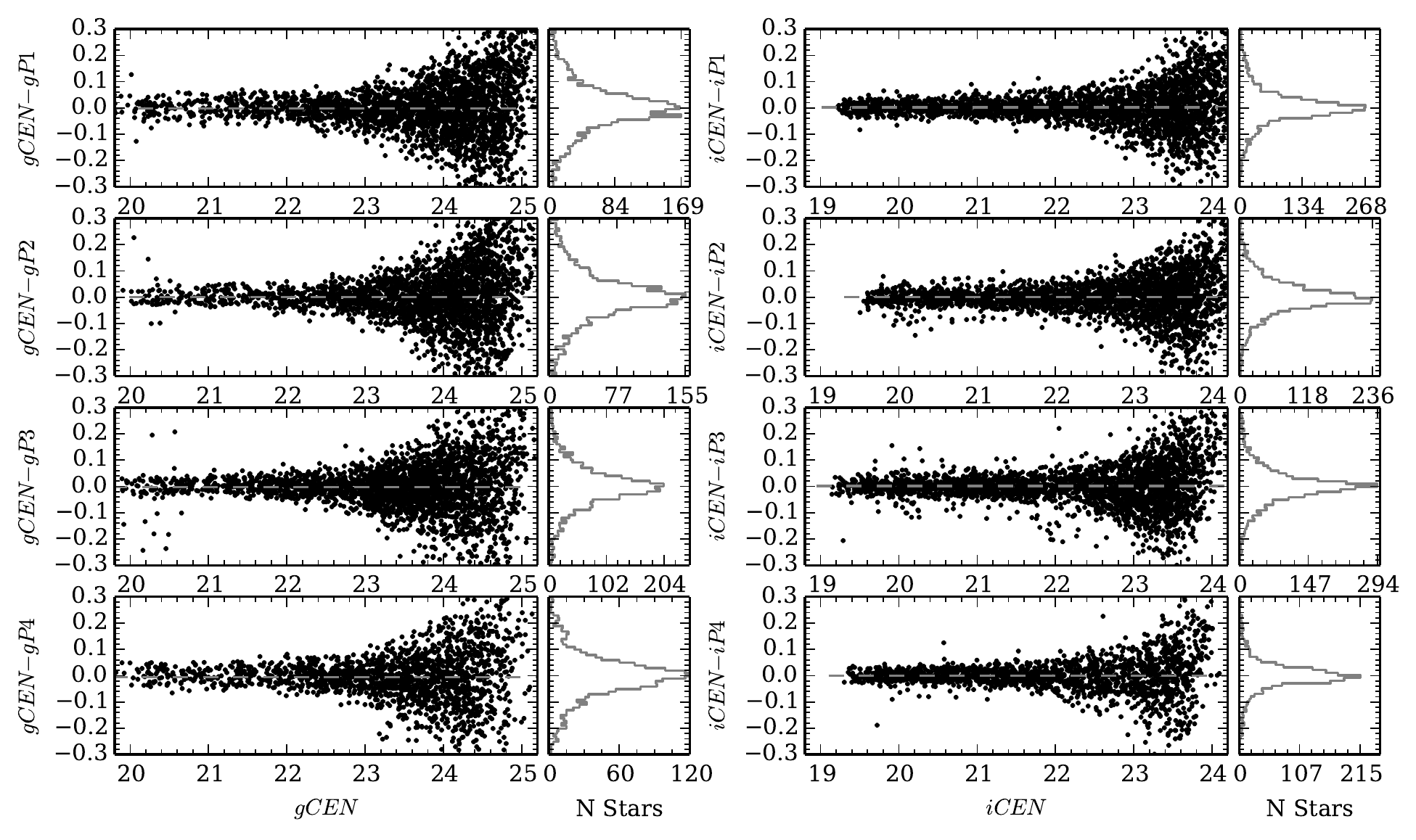}
\caption{Demonstration of the calibration check between fields, using overlapping regions.  Each plot shows the difference in measured magnitude for each star in the overlapping regions between the central (CEN) and outer (P1-P4) fields. The grey dashed line represents the median difference in each and is expected to be close to zero. A histogram shows the frequency distribution for each overlap region.  Figures on the left represent $g$, while figures on the right represent $i$.}
\label{fig:calcheck}
\end{figure*}

Once each catalogue had been calibrated to the SDSS photometric system, the outer fields were calibrated to the central field using stars in the overlapping regions between fields.  The stars in each of the overlapping regions have been extracted and measured separately for each image, and thus provide two measurements which can be used for calibration.  This was performed in the same manner as the SDSS calibration, with the central field serving as the reference to calibrate from.  The stars in each of the overlapping regions were matched according to their sky coordinates, and each outer field then calibrated to the central field in each filter.  A check of the calibrations was then performed for each field as follows.  Again using the overlapping regions, the magnitude difference between measurements of stars in the outer field and the central field were compared in each filter as a function of magnitudes measured in the central field.  The frequency distribution that resulted was also considered.  A consistent calibration would show a small scatter and no systematic offset.  Some inconsistencies were noticed, particularly in the $i$-band images with poorest seeing. Closer investigation revealed that the anomalous measurements came from stars which lacked coverage by one or more images in the co-added stack. These stars were removed from the catalogue using the exposure maps generated by the DECam community pipeline. To ensure accurate photometry, only stars with coverage from all three images in the stack for both filters were retained. The calibration process was repeated and checked again. The plots in Figure \ref{fig:calcheck} illustrate the calibration check, and show, as expected, that the magnitude difference is distributed about zero in each field (and filter) and flares out as magnitudes become fainter and measurement uncertainties increase.

Once satisfied that each of the outer fields had been accurately calibrated to the centre, a correction for Galactic extinction was applied.  The large DECam field of view affords noticeable variation in Galactic extinction.  The greatest variation is seen in the $g$-band, being $0.092<A_g<0.250$ (a differential variation of approximately 8\% in magnitude). In order to correct for this variation, and given the photometry has been calibrated in the SDSS photometric system, the extinction values used by SDSS in the direction of Sextans were adopted for our data set.  Although newer coefficients exist \citep{2010ApJ...725.1175S,2011ApJ...737..103S}, the coefficients used to obtain the SDSS corrections were $A_g=3.793E(B-V)$ and $A_i=2.086E(B-V)$ \citep{2002AJ....123..485S}, based on the extinction maps of \citet{1998ApJ...500..525S}.  Each star in each of our catalogues was matched to its nearest SDSS neighbour, and the corresponding extinction value applied.

\begin{figure}
\centering
\includegraphics[width=\columnwidth]{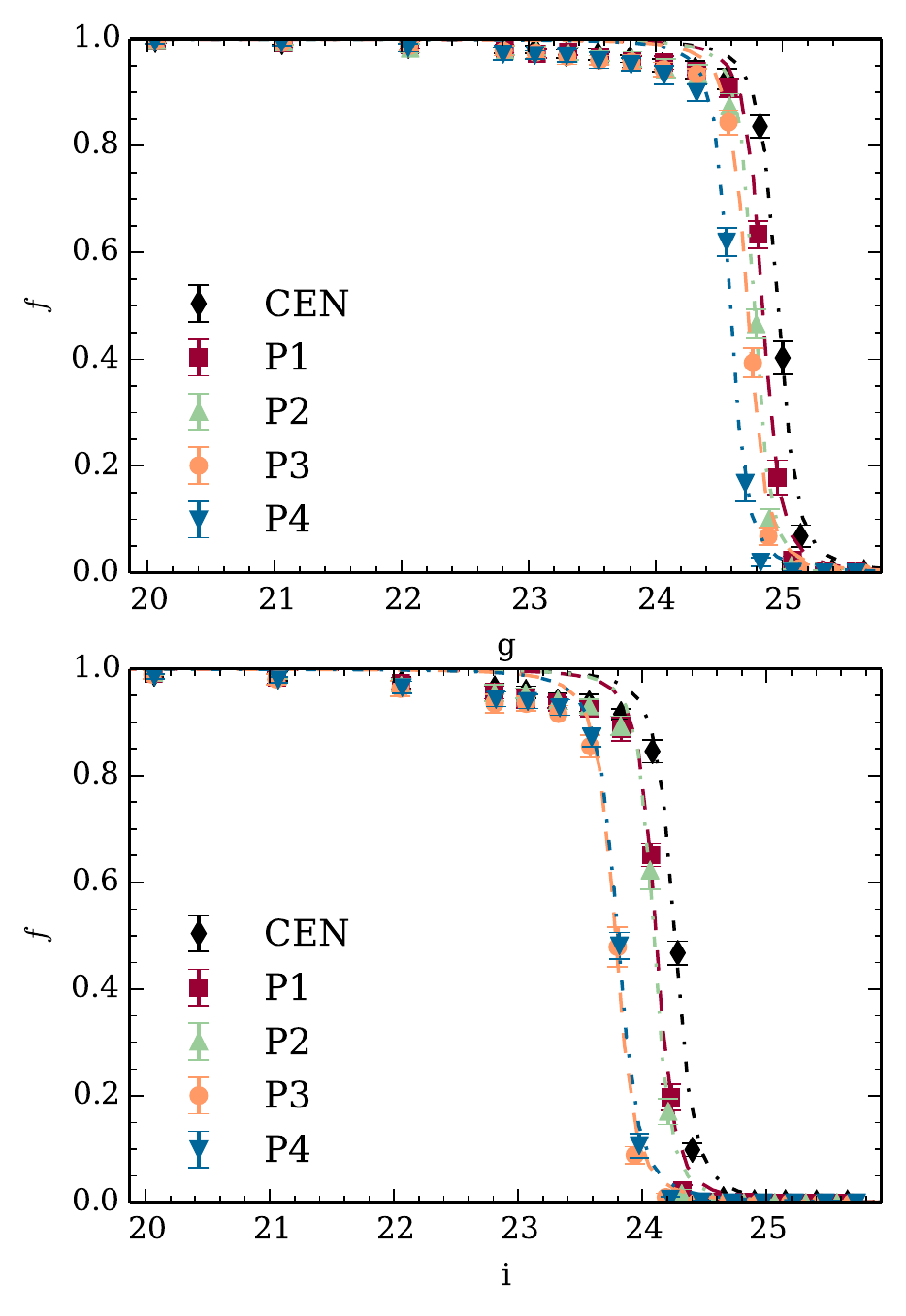}
\caption{Results of the photometric completeness test in $g$ (top) and $i$ (bottom) for each field.  Dashed lines represent the analytical relation described by \citet{1995AJ....109.1044F}. The 50\% and 90\% completeness limits are given in Table \ref{tab:data}.}
\label{fig:completeness}
\end{figure}

The final step in the photometric process was to perform completeness tests.  These were carried out separately in each filter for each field, in order to determine photometric depth and accuracy.  The IRAF \textit{addstar} task was used to add artificial stars to each image in random R.A. and Dec. Stars were added in magnitude bins, 1\,mag apart at the bright end and decreasing to 0.25\, mag apart at the faint end in the range between 20 and 26 mag, to best sample the completeness drop-off.  The image was then run through the photometry pipeline and the artificial stars recovered.  This process was completed 15 times for each image to build up a statistically robust sample, with $\simeq58,000$ stars added in total to each image in each filter.  The 50\% completeness level was determined by fitting the relation described by \citet{1995AJ....109.1044F} to the fraction of recovered stars.  The results of the completeness test for each field are shown in Figure \ref{fig:completeness} and summarised in Table \ref{tab:data}.  The completeness across all fields is relatively consistent, with the exception of P3 and P4 in the $i$-band, which are approximately 0.5 mag shallower than the deepest field (CEN) due to slightly poorer seeing.
 
Photometric accuracy was determined by comparing the input magnitude to the measured magnitude of each artificial star recovered.  An exponential function was fit to these results and used as a reference for determining the uncertainty of each individual star in each catalogue.  This was performed separately for each image, in each photometric band.  

\section{Foreground Discrimination}
\label{sec:fgd}
Clean discrimination between stars belonging to the Sextans dwarf and stars belonging to the MW halo is of paramount importance for the identification and characterisation of extended stellar substructure associated with the dwarf. Selection of stars based on their position in the colour-magnitude diagram (CMD) aids greatly in achieving such discrimination. This forms the basis for a technique often referred to as matched filtering. In this context, the colour-magnitude information of a stellar catalogue is used to create a filter for investigating its spatial distribution, and has proved to be a highly successful means for investigating stellar substructure in a low signal to noise environment \citep[e.g.][]{2002AJ....124..349R,2001ApJ...548L.165O,2014MNRAS.444.3139M,2015ApJ...803...80K}. Adopting this philosophy, we created a weight mask for our stellar catalogue, to assign each star a weight based on its position in the CMD.  

We followed the methodology used by \citet{Roderick:2015jj} for creating a graduated mask based on an isochrone that best describes the stellar population. However, rather than using a model isochrone, we determined the locus of the Sextans stellar population empirically.  This approach was possible because Sextans has a much more densely populated CMD than that of Hercules analysed by \cite{Roderick:2015jj}. The empirical fit was determined using a CMD of stars inside the half-light radius of Sextans, but having had the horizontal branch removed to ensure that only the main sequence, sub-giant and giant branch were present.  The horizontal branch and blue straggler populations contain fewer stars and will be used during the analysis for a consistency check. Consequently they should not be used to identify potential over-densities.  The remaining CMD was binned down the magnitude axis ($g$), in increments of 0.002, and the median colour-value determined for each bin.  Thus, for each bin in $g$, the colour ($g-i$) corresponding to the most well populated part of the CMD was known. These values were then used as the basis for a second binning process with larger bin size in $g$, and the corresponding colour values determined for the new bins. This iterative process was continued with increasing bin size, until the ridge line of the Sextans isochrone emerged.  The final isochrone model was obtained by performing a cubic interpolation over the result of the iterative process.

\begin{figure*}
\centering
\includegraphics[width=\textwidth]{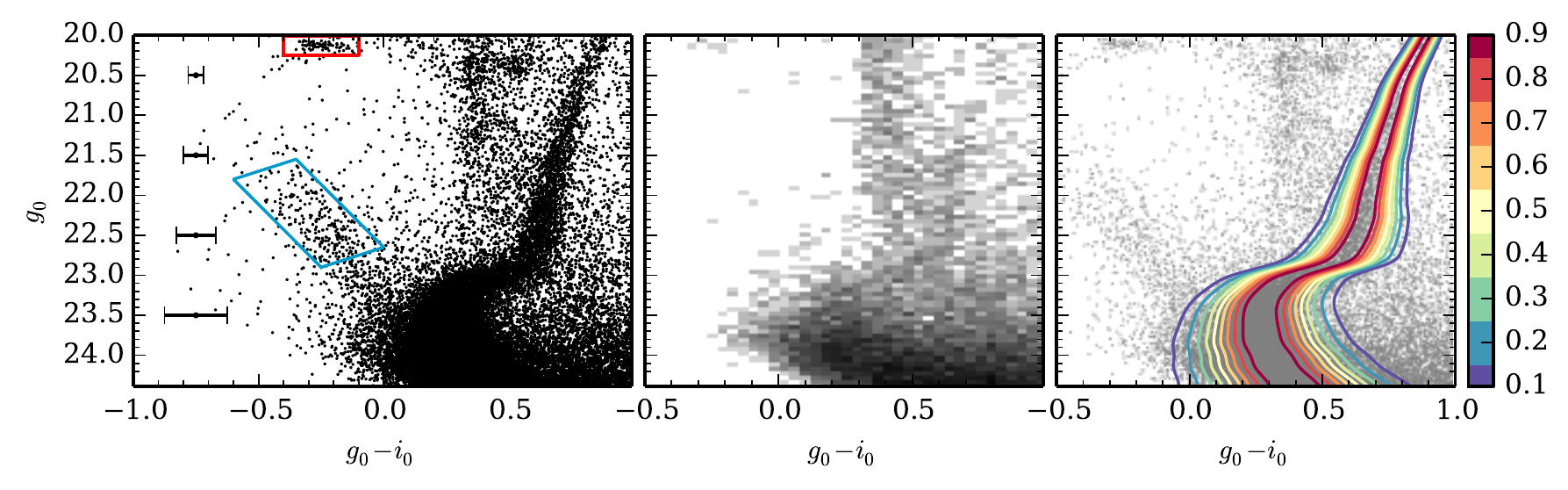}
\caption{Left: CMD of all stars within the central half-light radius of Sextans.  Blue and red boxes denote regions of candidate BSS and BHB stars respectively. Centre: 2D histogram of the complete Sextans catalogue, showing considerable contamination from MW halo stars. Right: Contour plot of the weight mask created from an empirically derived model of the Sextans isochrone. Mask width increases with photometric uncertainty. The colour bar represents the contour values.} 
\label{fig:mask}
\end{figure*}

Once the empirical model had been determined, it was used to create a mask for weighting stars in each of the five catalogues of our fields, similar to that of \citet{Roderick:2015jj}.  The mask was created by first binning the isochrone in 0.02 mag increments down the $g$ magnitude axis. A Gaussian profile along the colour range was calculated for each bin, centred on the Sextans stellar locus given by the empirical model.  The width of each Gaussian was determined by the photometric uncertainty of $g-i$, and the amplitude set to 1. Thus stars could be assigned a weight, $w$, between 0 and 1 according their proximity to the stellar locus. The mask for the central field is shown in the right panel of Figure \ref{fig:mask}.  The shape of the mask reflects the photometric uncertainties, and is based on the stellar population within the half-light radius of Sextans. Note that the areas of the CMD corresponding to potential horizontal branch or blue straggler stars have been been assigned `nan' values to ensure they are omitted from the over-density identification process to follow.

A separate mask was made for each field using the same empirical isochrone model. However the width of the Gaussian profile calculated along the colour range for each mask varied according to the photometric uncertainties in each individual field. This allowed the weighting to reflect the varying uncertainties in the photometry caused by the variation in seeing between the different images.  Each mask was applied to its corresponding catalogue, with stars weighted according to their proximity to the ridge line of the empirical isochrone as described above.  

Once the membership weights were applied, the five separate catalogues were merged into a single catalogue.  Where there were duplicate measurements of stars from the overlapping regions, the measurement from the image with the better seeing was retained, and the other discarded.  Finally, the combined catalogue was cut at the 90\% completeness level of the shallowest field ($g=24.39$), to ensure even depth across the entire observed field-of-view.  This final catalogue contained more than 46,000 stars, reaching a photometric depth approximately 1 mag below the main sequence turn off.

\section{Spatial Mapping}
\label{sec:detect}
Having weighted each star in the catalogue according to colour and magnitude, we used the weights to create a map of the spatial distribution of Sextans stars. The weights were used to subdivide the catalogue into two groups, a foreground catalogue and a Sextans member catalogue, similar to the method used by \citet{Roderick:2015jj}. This then provides a catalogue which can be used to create a map of the foreground which can then be subtracted from the map of Sextans members to reveal substructure in the surrounding region, similar to \citet{2015ApJ...803...80K}.

\begin{figure*}
\centering
\includegraphics[width=\textwidth]{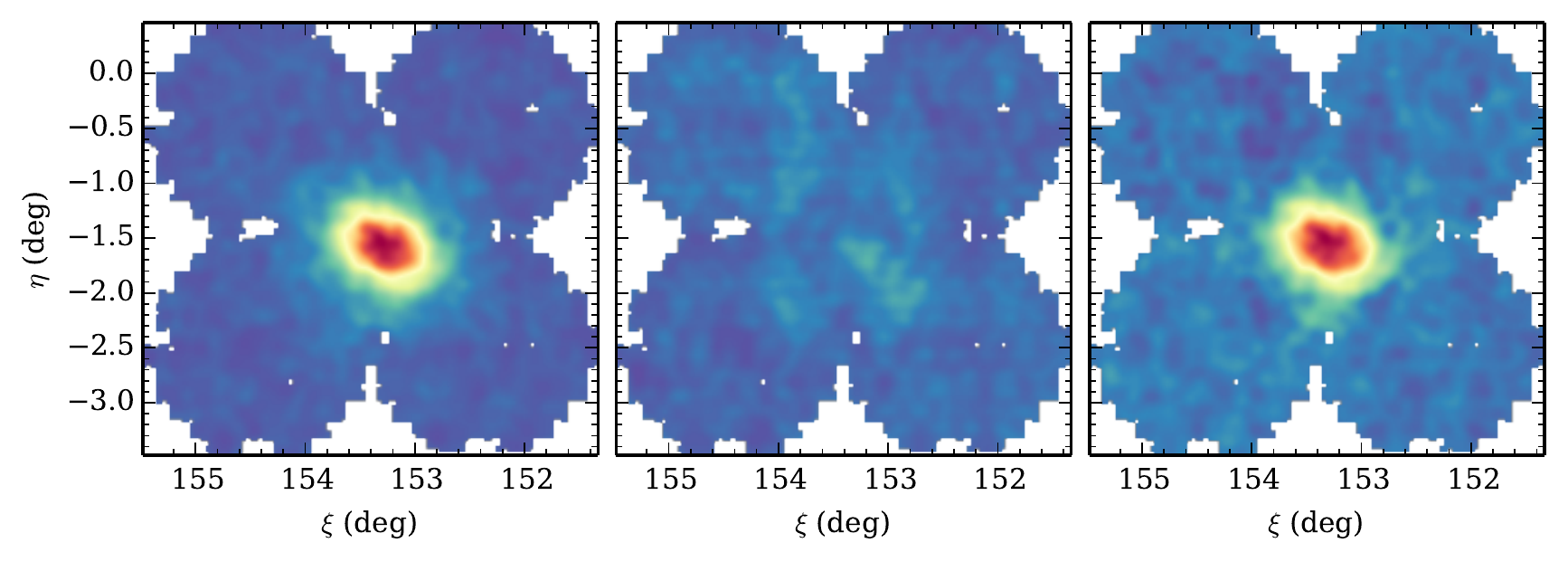}
\caption{Each step in the process of creating the detection map. Left: smoothed 2D histogram of Sextans members. Centre: smoothed 2D histogram of `foreground'. Right: final flat fielded, foreground subtracted detection map. The colour scale increases from blue to red, and is the same for left and centre panels, but differs in the right panel to make the features of each visually clear.}
\label{fig:histograms}
\end{figure*}

To create the maps, each catalogue was binned into a normalised 2D histogram of R.A. and Dec. with equal bin sizes of $30\arcsec\times30\arcsec$, which we will refer to as pixels. Each histogram was smoothed with a Gaussian kernel, to create a density map.  The normalised foreground map (centre panel of Figure \ref{fig:histograms}) was then used as a form of flat field (such that the average pixel value across the field was equal to one), and divided through the Sextans map (left panel of Figure \ref{fig:histograms}).  Finally, the foreground map was subtracted from this flat-fielded map in order to reveal substructures present in the field (right panel of Figure \ref{fig:histograms}). We note that, as a result of the large smoothing kernel used, this final `foreground' corrected image is free from the pattern created by the inter-chip gaps between the CCDs.

Several different cut-off weights were tested in order to find the optimal value for our density maps.  Varying the cut-off weight effectively changes the width of the selection region around the empirical isochrone, including more or less of the Sextans stellar population.  A weight $w = 0.3$ was determined as the optimal cut-off point since this encompasses the body of Sextans stars in the CMD, without cutting too many away or including a large fraction of Galactic foreground stars (refer to right-hand panel in Figure \ref{fig:mask}).

Varying the size of the Gaussian kernel changes the scale of the detected over-densities.  In order to find an appropriate smoothing factor, kernels of 3, 5, and 7 pixels (corresponding to $90\arcsec, 150\arcsec$ and $210\arcsec$), were tested.  A kernel smoothing of 7 pixels was chosen since it provided the most detail with the least amount of noise in the detections.  While the 3, and 5 pixel kernels revealed similar over-all structures, they also provided a large amount of `noisy' detections.  The detection maps shown in Figure \ref{fig:histograms} demonstrate the 7 pixel ($210\arcsec$) smoothing kernel.

\begin{figure*}
\centering
\includegraphics[width=0.7\textwidth]{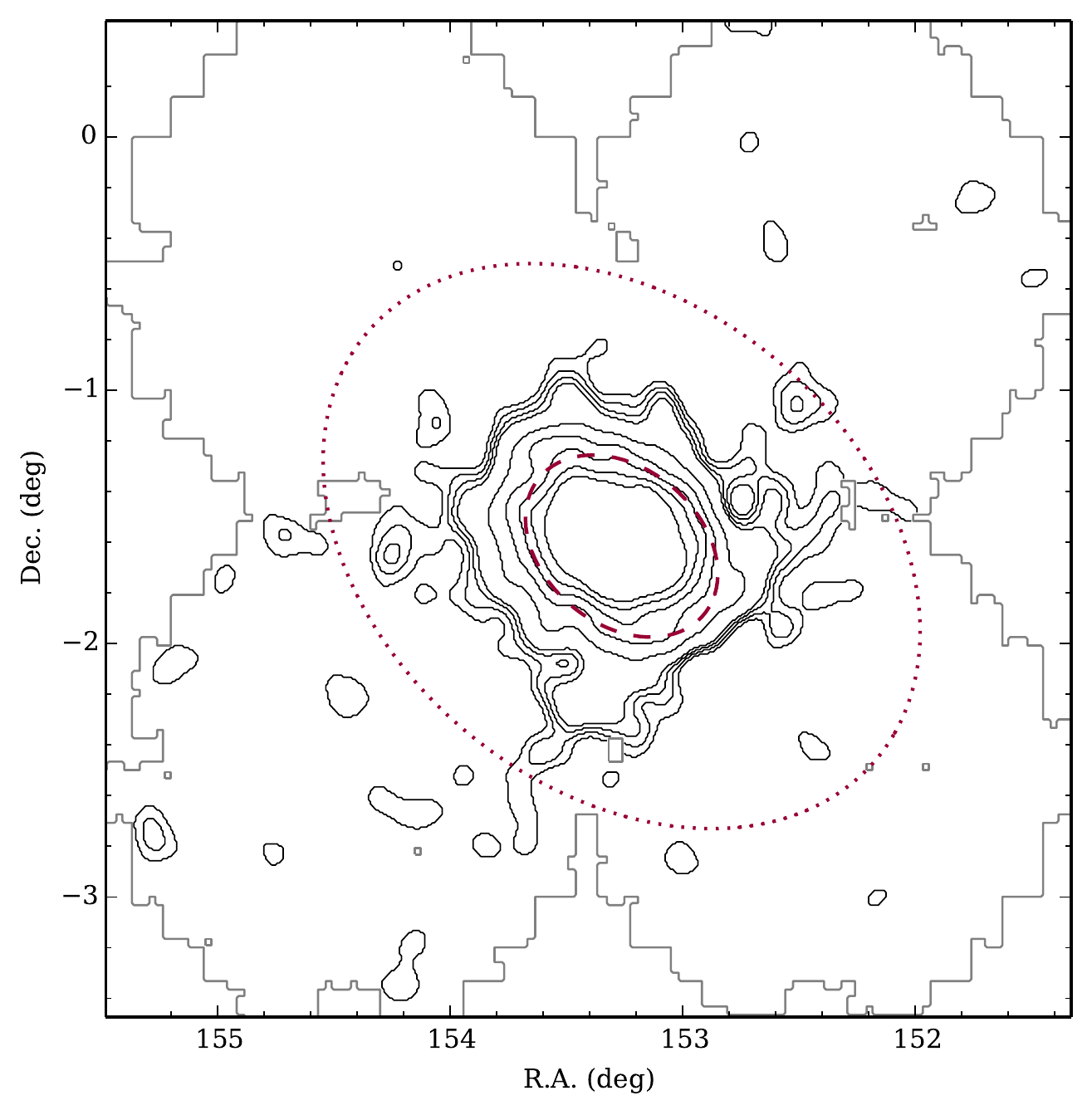}
\caption{Contour map of the foreground subtracted spatial distribution of Sextans. Contours represent 2,3,4,5,10,15,20,25,30 times $\sigma_t$ above the mean pixel value, where the mean pixel value is determined from the region outside three times the half-light radius of Sextans and taken to represent the average field value. Red dashed and dotted lines represent the core and tidal radii respectively.}
\label{fig:contours}
\end{figure*}

Finally, with the detection of over-densities in the outskirts of the galaxy in mind, we define a detection threshold in terms of the mean pixel value in the outer regions of the smoothed, foreground subtracted map. Since the signal from the centre of Sextans is so strong, we consider all pixels in the map outside of three times the half-light radius \citep[$27\farcm8$,][]{1995MNRAS.277.1354I}, corresponding approximately to the furthermost half of each of the outer four fields. We then define contours in terms of the standard deviation, $\sigma_t$ (where $_t$ is a reminder that this is used to determine the detection threshold), above the mean pixel value, with the lowest threshold corresponding to $2\sigma_t$ above the mean. Figure \ref{fig:contours} shows the full map with contours at 2,3,4,5,10,15,20,25,30 times $\sigma_t$.

\section{Structural Parameters}
\label{sec:profile}
Using the spatial map created in the previous section, we determined the orientation angle, $\theta$, and ellipticity, $\epsilon$, of Sextans.

We determined $\theta$, and $\epsilon$ for each of the contours in our spatial map, averaging the results and using the variation between measurements to estimate more robust uncertainties, similarly to \citet{McConnachie:2006gl}. This takes into account the variation in shape and orientation with radius of the galaxy, giving a more independent measurement of the parameters. A maximum-likelihood bivariate-normal-fit from the astroML\footnote{http://www.astroml.org/} \citep{2012cidu.conf...47V} machine learning package was used to determine $\theta$, and $\epsilon$, as well as the centroid for the spatial distribution in each case. The results of our analysis show $\theta = 57.5^\circ\pm5.3^\circ$, $\epsilon = 0.29\pm0.03$ and a centroid offset in degrees from the central R.A. and Dec of $0.014\pm0.003$ and $0.002\pm0.008$ respectively. Our results for $\theta$ and $\epsilon$, and the centre of Sextans, are consistent with the values from \citet[$\theta=56^\circ\pm5^\circ$, $\epsilon=0.35\pm0.05$,][]{1995MNRAS.277.1354I}.

\begin{figure*}
\centering
\includegraphics[width=\textwidth]{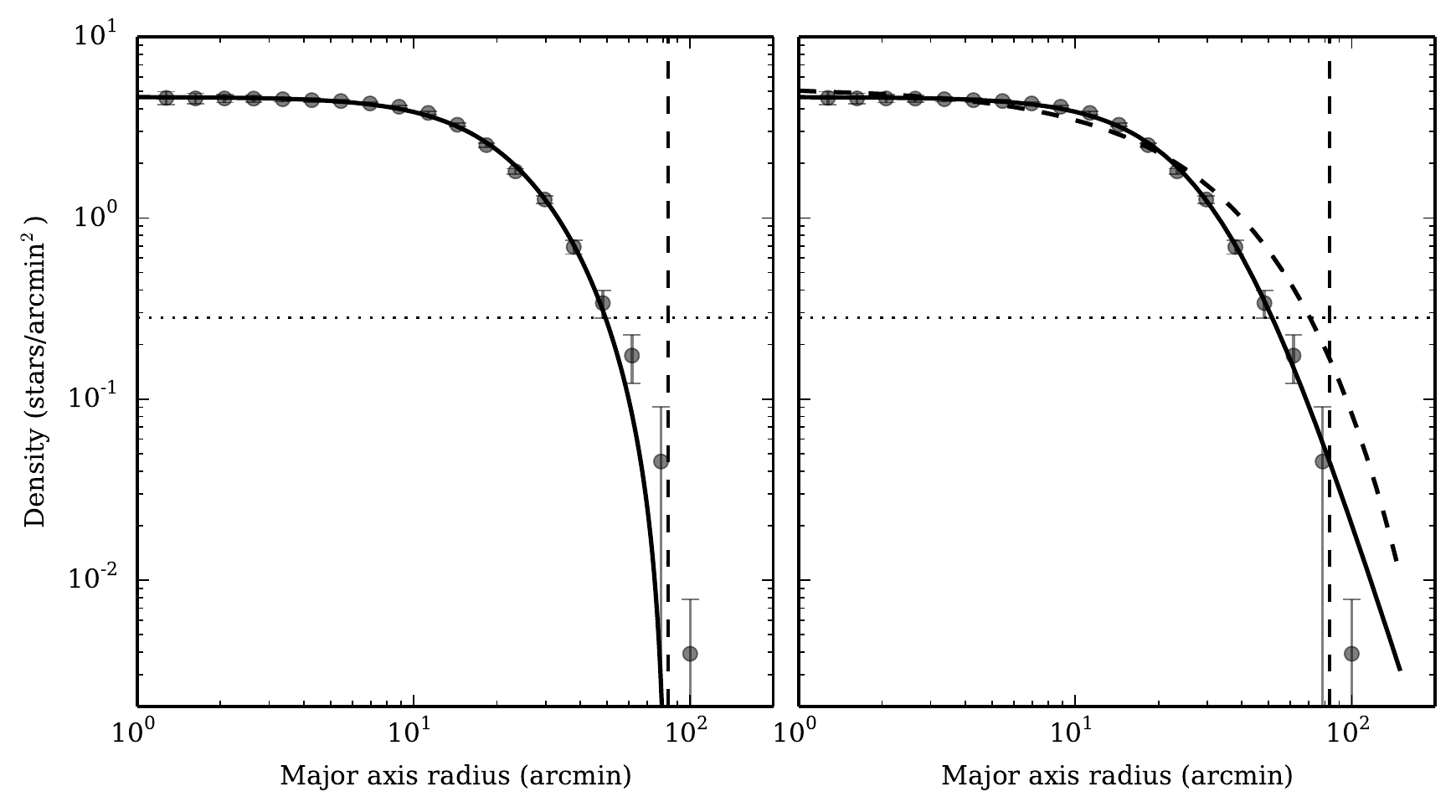}
\caption{Radial profile of Sextans showing number density of stars vs. major axis distance from centre. Markers show the profile using our own values for $\theta$ and $\epsilon$ determined by the astroML machine learning package. Error bars are given by Poisson statistics.  The dashed line in each plot indicates the tidal radius determined by fitting a King profile.  The dotted line in each plot shows the average background density. Left: Solid black line shows the King profile fit.  Right: Dashed black line shows the exponential profile fit, solid black line shows the Plummer profile. Tidal and core radii determined from fitting the King profile are given by $r_t = 83.2\arcmin\pm7.1\arcmin$ and $r_c = 26.8\arcmin\pm1.2\arcmin$ respectively. The core radius determined by the Plummer profile is given by $r_c = 23.0\arcmin\pm0.4\arcmin$}
\label{fig:profile}
\end{figure*}

We also produced a radial profile based on the foreground subtracted map. Using the values determined in the previous paragraph, we placed logarithmically spaced, concentric, elliptical annuli about the centre of Sextans out to a major axis distance of $100\arcmin$. This major axis length was chosen since it encompasses as much of the data set as possible, without meeting the edge of our fields. The stellar number density was noted for each annulus. The results are shown in Figure \ref{fig:profile}, with the error-bars reflecting the measurement uncertainties in the counts, based on Poisson statistics, as well as the uncertainty in the foreground map. A King profile \citep{1962AJ.....67..471K} was fit to these measurements, as well as an exponential profile, similar to \citet{1995MNRAS.277.1354I}.  Figure \ref{fig:profile} shows both profiles, the King profile (left), and the exponential profile (right). We note that although \citet{1995MNRAS.277.1354I} found a similarly good fit for the exponential profile as the King profile, we see a better fit for the King profile than the exponential.Consequently, we also fit a Plummer profile \citep{1911MNRAS..71..460P}, also shown in the right panel of Figure \ref{fig:profile}, which shows an improved fit.  The fit for the King profile yielded a core radius of $r_c = 26.8\arcmin\pm1.2\arcmin$, and a tidal radius of $r_t = 83.2\arcmin\pm7.1\arcmin$, compared to \citet[$r_c=16.6\arcmin\pm1.2\arcmin$, $r_t=160\arcmin\pm50\arcmin$,][]{1995MNRAS.277.1354I}, and \citet[$r_c=15\arcmin$, $r_t=90\arcmin$,][]{1990MNRAS.244P..16I}.  Comparatively, the fit for the Plummer profile yielded a core radius of $r_c = 23.0\arcmin\pm0.4\arcmin$. Although the agreement between the data and the best fitting King model is not particularly good at large radii, there is only tentative evidence for a break radius to indicate extra-tidal features. This will be discussed further in Section \ref{sec:disc}.

\section{Substructure Analysis}
\label{sec:identify}
We follow the methodology of \citet{Roderick:2015jj} to conduct our substructure analysis. The procedure is similar in nature to that of \citet{2015ApJ...803...80K}, however rather than determining a significance statistic for each pixel in our foreground subtracted map and then searching for substructure, we use the contour levels across our full map and assess the significance of each structure after it is identified.

We use the Python package \textit{scipy.ndimage}\footnote{http://docs.scipy.org/doc/scipy/reference/ndimage.html} to identify and label each individual over-density.  This package looks for groups of adjoining pixels in an image, and labels each group as a different segment.  The segmentation process was performed on each of the five lowest thresholds used to create the contour map in Section \ref{sec:detect} ($ 2\sigma_t, 3\sigma_t, 4\sigma_t, 5\sigma_t$ and $10\sigma_t$).  We perform the analysis on multiple thresholds in order to obtain a better understanding of the significance of the outer structure of Sextans on different levels.

Before assessing the significance of each segment, or over-density, a control region was determined for testing and calibrating against.  Ideally, the significance of an over-density is determined by comparing it to a sample of the foreground.  This is difficult since the body of Sextans takes up a large portion of our field-of-view. During the detection of over-densities we consider anything above the detection threshold as being a potential over-density, we therefore consider anything below the detection threshold as part of the foreground.  Thus, we define a control region for testing by taking the region of sky below the detection threshold. Having determined the segmentation of the detected over-densities, as well as appropriate regions from which to take a control sample, all the corresponding stars from the catalogue were separated into their relevant groups for the significance testing.  When separated into their relevant groups, stars from the complete catalogue were included regardless of their weight.  This was to ensure an indiscriminate analysis of the significance of each detection.

We determine a value for the significance of each over-density, $\zeta$, in the manner described by \citet{Roderick:2015jj}.  This $\zeta$ value is obtained by counting how many stars are in close proximity to the isochrone for each over-density, compared to the control region.  A star with close proximity to the isochrone is defined as having $w\geq0.9$.  The number of stars in an over-density that fit this criteria is given by the quantity $N_{w\geq0.9}(OD)$.  Note that since the main body of Sextans forms part of this analysis, we have excluded stars inside the region defined by the $15\sigma_t$ threshold. This was done in order to obtain a significance value that reflected the star content in the extremities of Sextans, rather than the centre. For each over-density, the control region was randomly sampled 10,000 times, selecting a number of stars equal to the over-density being tested and counting those with $w\geq0.9$.  A frequency distribution was built from these values, and a Gaussian fit to determine the mean, $\langle N_{w\geq0.9}(CS)\rangle$, and standard deviation, $\sigma$.  The significance, $\zeta$,  was then determined as the distance of $N_{w\geq0.9}(OD)$ from $\langle N_{w\geq0.9}(CS)\rangle$ in units of the standard deviation:

$$
\zeta = \frac{N_{w\geq0.9}(OD) - \langle N_{w\geq0.9}(CS)\rangle}{\sigma}
$$

\begin{figure}
\centering
\includegraphics[width=\columnwidth]{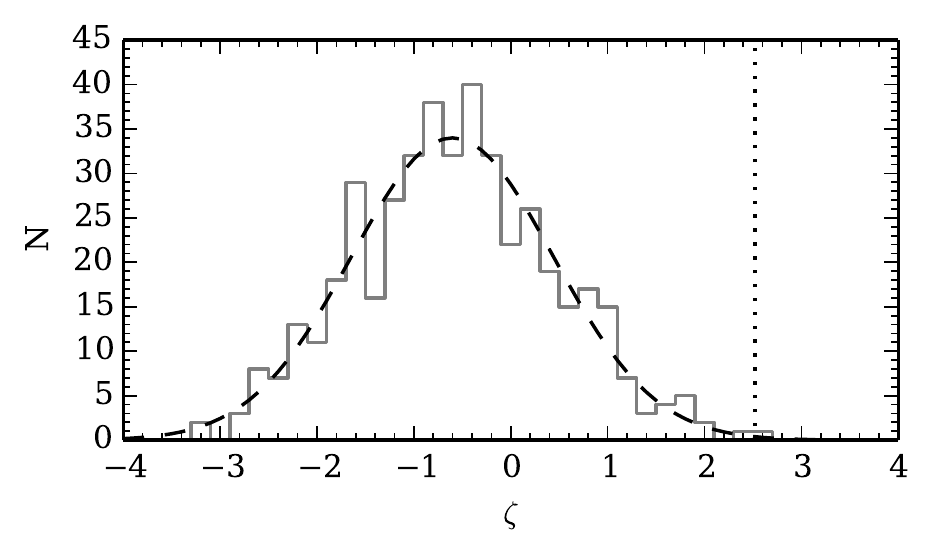}
\caption{Frequency distribution of $\zeta$ values obtained testing the null hypothesis (grey line). A Gaussian model (dashed line) yields a mean of -0.6, and a standard deviation of 1.04. Values above three standard deviations from the mean are considered statistically significant (dotted line).}
\label{fig:null}
\end{figure}

\begin{figure}
\centering
\includegraphics[width=\columnwidth]{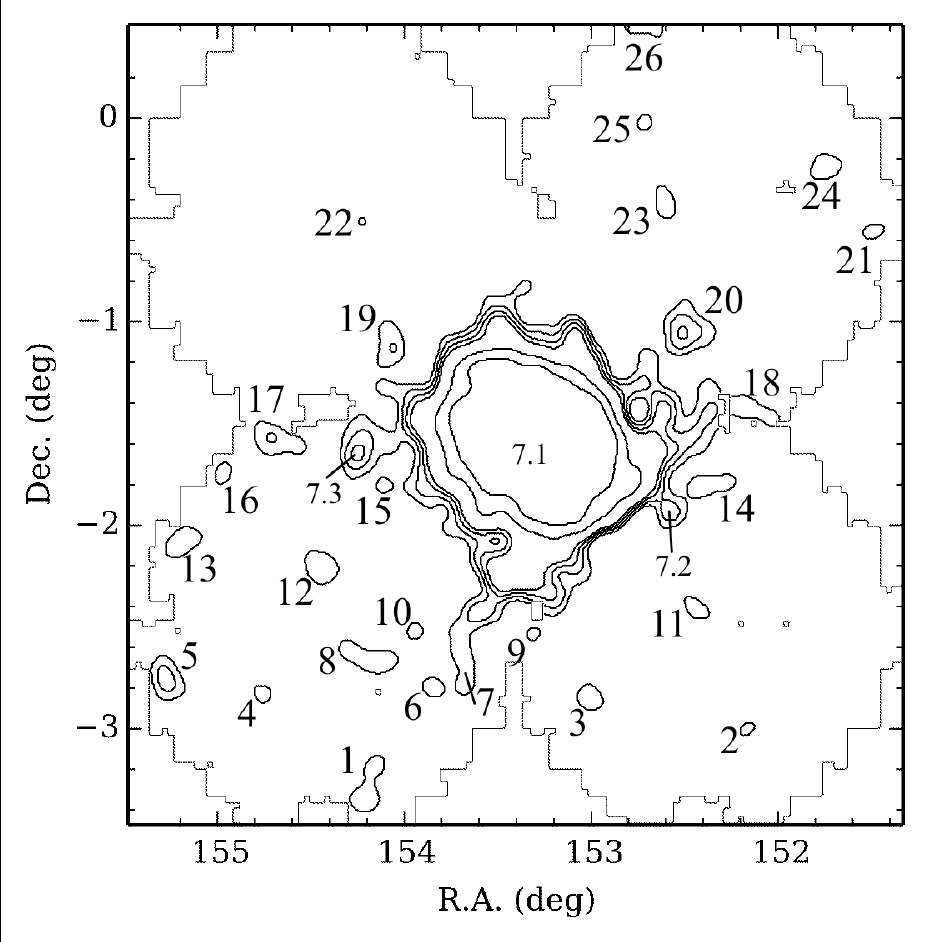}
\caption{Labelling convention used for over-densities discussed in the text. Labels also correspond to Table \ref{tab:zeta} and Figures \ref{fig:zeta} and \ref{fig:cmds}.}
\label{fig:labelling}
\end{figure}

\begin{figure*}
\centering
\includegraphics[width=\textwidth]{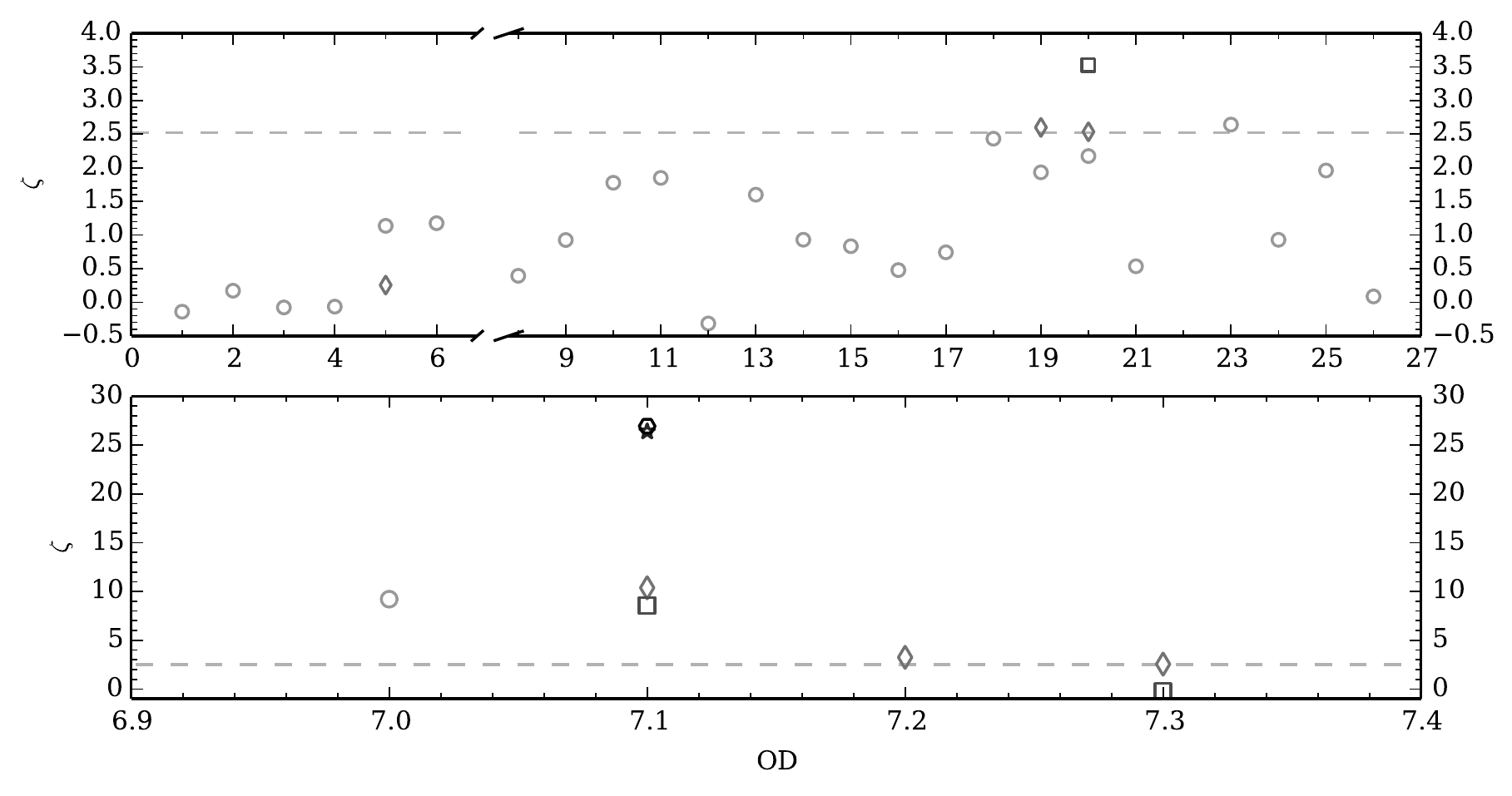}
\caption{The significance value ($\zeta$) for each segment, or over-density identified by the segmentation process.  Note the bottom panel is a close up of the corresponding region in the top panel (for visual clarity). Circles, diamonds, squares and stars and hexagons represent thresholds $2\sigma_t$ to $10\sigma_t$ respectively. The grey dashed line is at $\zeta=2.52$, above which detections are considered significant. }
\label{fig:zeta}
\end{figure*}

In order to understand the confidence of our $\zeta$ values, we performed a test of the null hypothesis. Our significance test was performed on 400 randomly chosen segments (varying in size from  4 to 360 square arcminutes) created from the control region.  Since these segments are dominated by foreground, the expectation is that a frequency distribution of the $\zeta$ values obtained should be centred approximately at $\zeta=0$.  Figure \ref{fig:null} shows this distribution.  A Gaussian function modelling this distribution yields a mean of -0.6, and a standard deviation of 1.04. Choosing significant values of $\zeta$ to be three standard deviations above the mean, we consider over-densities with $\zeta\geq2.52$ to be statistically significant. The results of this test for all the segments identified are shown in Figure \ref{fig:zeta}. The test results for those over-densities deemed significant are summarised in Table \ref{tab:zeta}. Since the segments at each high detection threshold are also found at the lower detection threshold, the labelling used at the lowest threshold ($2\sigma_t$) was adopted for all thresholds.  Where an over-density broke into multiple detections at a higher threshold, a decimal labelling system was used (e.g. OD 7 contains OD 7.1, OD 7.2 and OD 7.3). Figure \ref{fig:labelling} illustrates the labels discussed in the text.

\begin{table}
\begin{center}
\caption{Results for the over-densities with $\zeta\geq2.52$. Each horizontal line divides the group by detection threshold, with the top group found at $2\sigma_t$ and the bottom at $10\sigma_t$.}
\label{tab:zeta}
\begin{tabular}{lcrrrr}
\hline\hline
Segment &  &$N_{*}$ & $N_{w\geq0.9}(OD)$ & $\langle N_{w\geq0.9}(CS)\rangle$ & $\zeta$ \\
\cline{1-1}\cline{3-6}
OD 7.0 &&  1303 & 190 & 101 & 9.21  \\
OD 23.0 &&  14 & 4 & 1 & 2.64  \\
\hline
OD 7.1 &&  825 & 144 & 64 & 10.38  \\
OD 7.2 &&  38 & 9 & 2 & 3.25  \\
OD 7.3 &&  42 & 8 & 3 & 2.55  \\
OD 19 &&  4 & 2 & 0 & 2.60  \\
OD 20 &&  50 & 9 & 3 & 2.54  \\
\hline
OD 7.1 &&  549 & 95 & 42 & 8.56  \\
OD 20 &&  8 & 4 & 0 & 3.53  \\
\hline
OD 7.1 &&  1915 & 447 & 148 & 26.39  \\
\hline
OD 7.1 &&  1433 & 377 & 111 & 26.94  \\
\hline \hline
\end{tabular}
\end{center}
\end{table}

\begin{figure*}
\centering
\includegraphics[width=0.86\textwidth]{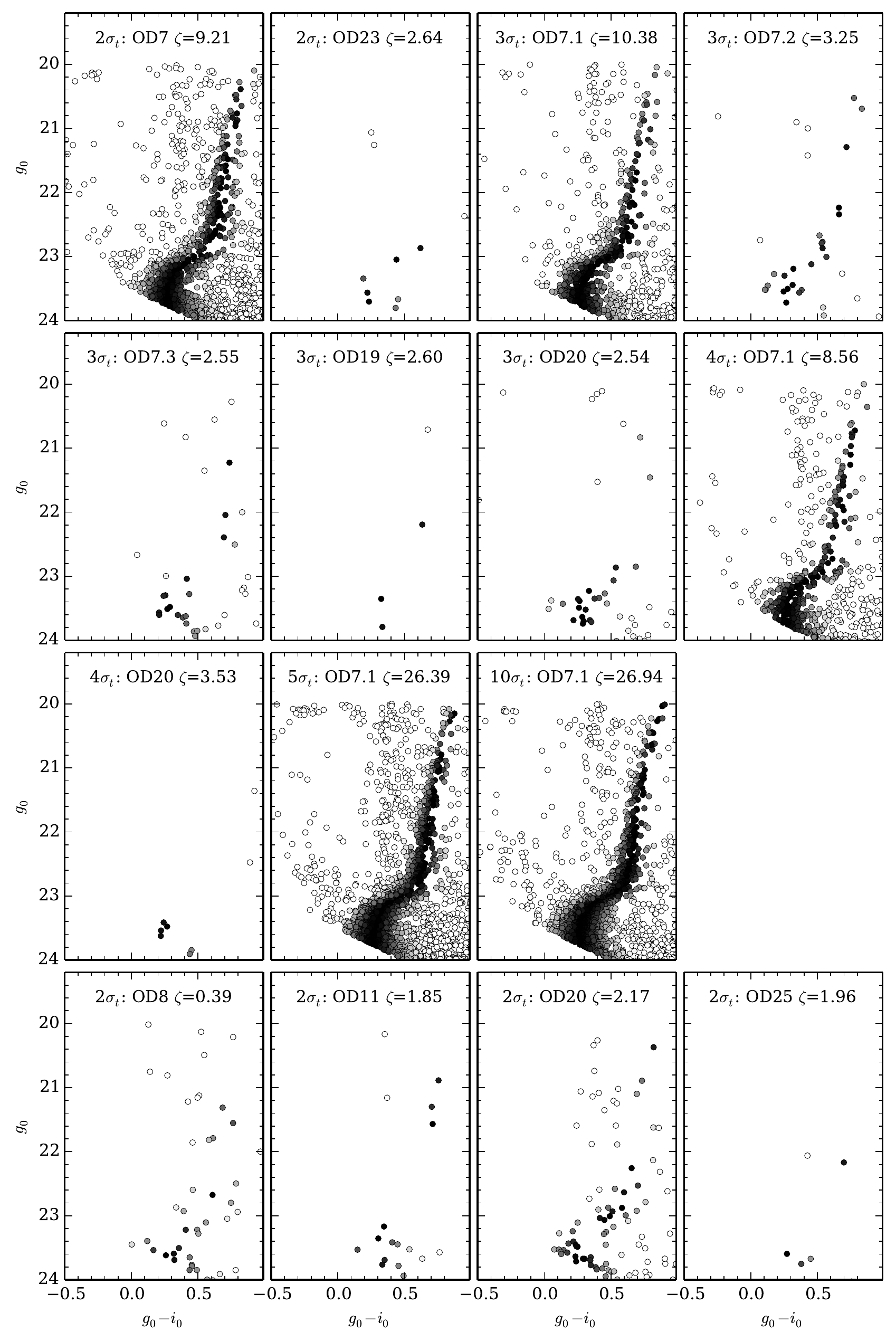}
\caption{Top three rows show CMDs for all detections with $\zeta\geq2.52$ (listed in Table \ref{tab:zeta}), bottom row shows examples of detections with $\zeta<2.52$. The detection threshold is labelled on each plot. Marker fill-colour correlates with mask weight (where 1 is black and 0 is white). Note horizontal branch features in OD 20, as well as the main region surrounding Sextans (OD 7, OD 7.1).}
\label{fig:cmds}
\end{figure*}

To further illustrate the significance of the detections with $\zeta\geq2.52$, colour-magnitude diagrams of over-densities with $\zeta\geq2.52$ have been included in Figure \ref{fig:cmds}.  A population of stars belonging to Sextans is evident in the main region surrounding the body of the dwarf (OD 7, OD 7.1 at the different detection thresholds). Horizontal branch and blue straggler stars are apparent in these CMDs as well. OD 20 also displays features of the Sextans population including horizontal branch stars (at the $3\sigma_t$ threshold). OD 7.2 and OD 7.3 and OD 19 contain fewer stars, however a statistically significant fraction of them are consistent with the Sextans empirical isochrone. The bottom row of CMDs in Figure \ref{fig:cmds} are from detections with $\zeta<2.52$. These diagrams show more obvious foreground stars relative to the number of stars with $w>0.9$.

As a final test of our detections, we investigated the distribution of background galaxies across the field. While apparent over-densities were detected, they were found to have no correspondence to any of the features discussed above. Furthermore, assessing their significance in the same way as our stellar over-densities yielded no significant $\zeta$ values, supporting the robustness of the star/galaxy separation and indicating that the substructure detected about Sextans is most certainly stellar.

\begin{figure*}
\centering
\includegraphics[width=\textwidth]{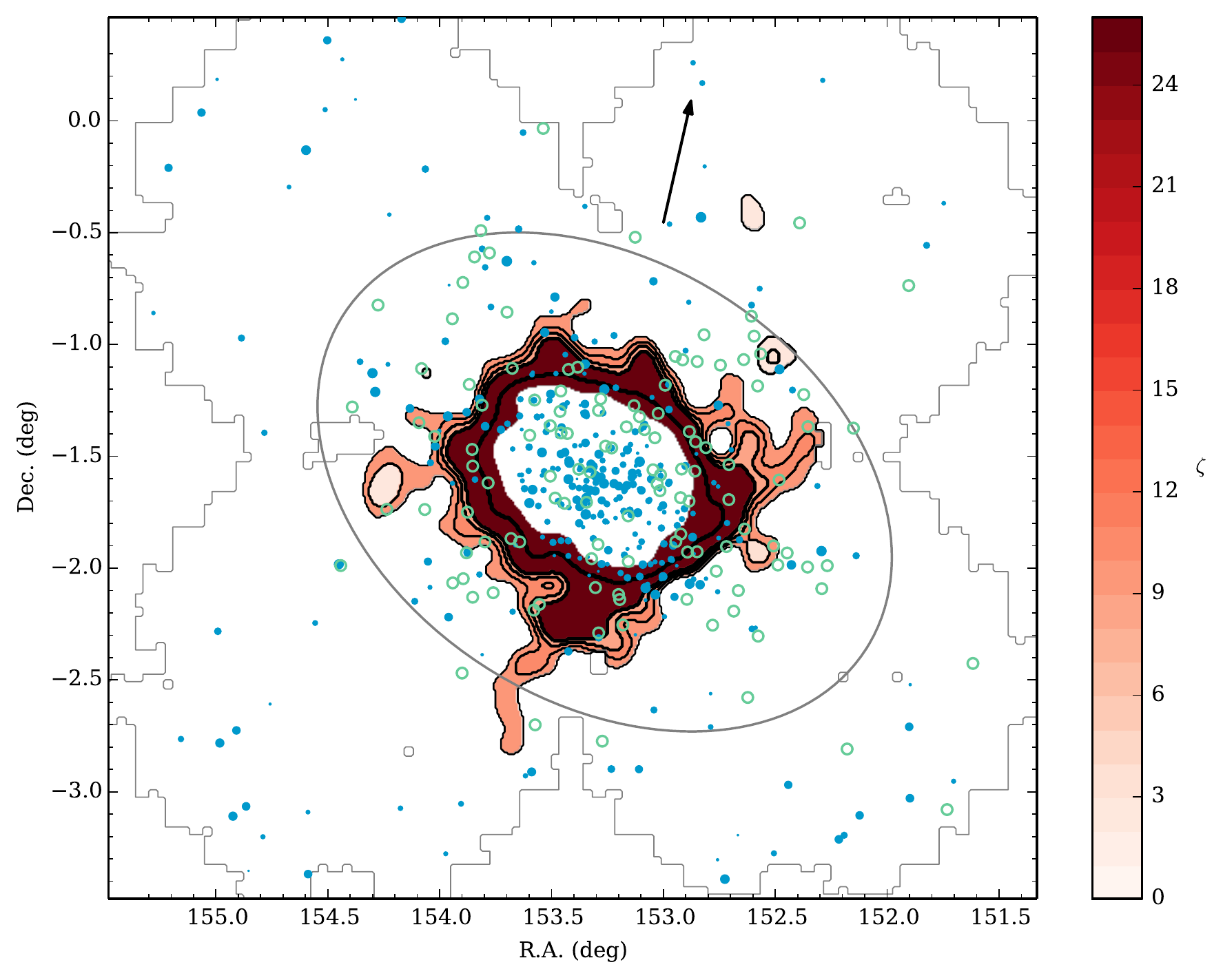}
\caption{This figure summarises the results of the stellar over-density significance testing across all detection thresholds.  The contours increase in line-width with an increase in detection threshold. All detection thresholds are over-laid on top of each other, and colour coded according to the $\zeta$ value.  Green open circles represent the Blue Horizontal Branch stars (outlined in red in Figure \ref{fig:mask}), while the blue filled circles represent Blue Straggler stars (outlined in blue in Figure \ref{fig:mask}).  Note that the size of the blue points is correlated with the brightness of these stars, where point size increases with brightness. The central region in white corresponds to the region above the $15\sigma_t$ threshold, where stars were not used to determine $\zeta$ due to the strong signal from Sextans. They grey ellipse represents the tidal radius determined from the King profile, and the black arrow shows the direction of Sextans' proper motion \citep{2008ApJ...688L..75W}.}
\label{fig:summary}
\end{figure*}

A summary of the stellar over-densities which were determined to be significant is shown spatially in Figure \ref{fig:summary}. This figure is discussed in more detail in Section \ref{sec:disc}. We also include a catalogue of all stars corresponding to the spatial distribution of over-densities in Figure \ref{fig:summary} as supplementary online data.

\subsection{The Centre of Sextans}
Although the focus of this paper is to look for substructure in the outskirts of Sextans, there have been interesting discussions on kinematic substructure more centrally located \citep[see][]{2004MNRAS.354L..66K,2006ApJ...642L..41W,2011MNRAS.411.1013B,2012ApJ...759..111K}.  Here we present a brief analysis of this inner region of our Sextans field.  Using our smoothed foreground subtracted map, we focus on the central region with contours representing 40, 50 and 60 times $\sigma_t$. It is important to note that this region will be more sensitive to any member stars that may have fallen into the foreground map, compared to the outskirts. However, the weight mask was applied carefully to minimise these effects. Also note that the contours do not represent the presence of the candidate BSS and HB star populations. These are investigated separately.

\begin{figure}
\centering
\includegraphics[width=\columnwidth]{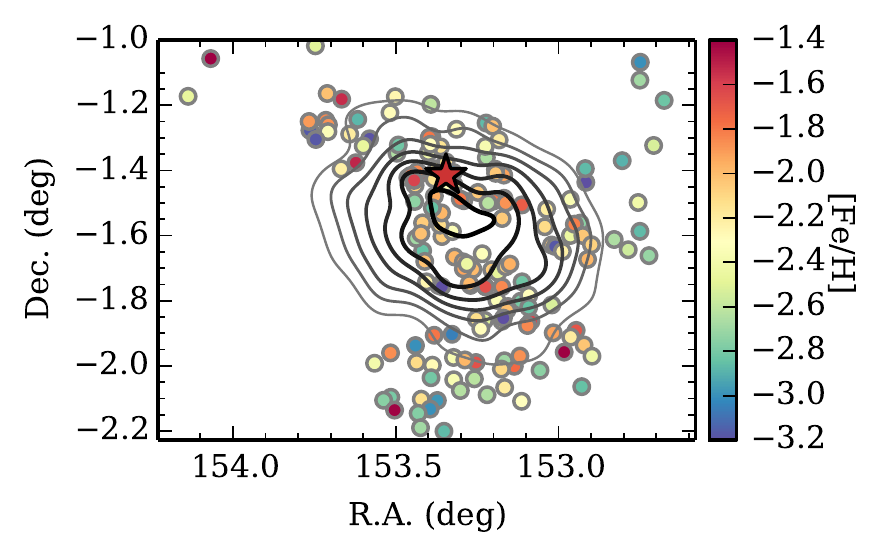}
\caption{Contour plot of central region of Sextans. Contours representing 15, 20, 25, 30, 35 and 40 times $\sigma_t$ are shown in increasing shades of grey to black.  A red star represents the location of a kinematically cold region detected by \citet{2006ApJ...642L..41W} at the 95\% confidence level. Coloured points represent red giant branch members identified by  \citet{2011MNRAS.411.1013B}, with the colour scale showing [Fe/H].  Note that the coverage by \citet{2011MNRAS.411.1013B} is not uniform, and therefore the lack of stars South-East of centre is a selection effect.}
\label{fig:centre}
\end{figure}

Figure \ref{fig:centre} displays the results of this analysis, with contours displayed at 20, 30, 40, 50 and 60 times $\sigma_t$ making comparison to the full map straight forward. The contours are shown in increasing shades of grey to black, and show that the densest part of Sextans is not actually at its centre, defined by  \citet{1995MNRAS.277.1354I}, but slightly toward the North-East of its centre. This is the same direction as the offset in coordinates found in Section \ref{sec:profile}. It is interesting to note that this is close to the kinematically cold region, marked with a red star in Figure \ref{fig:centre}, detected by \citet{2006ApJ...642L..41W}.  Figure \ref{fig:centre} also shows confirmed red giant branch stars from \citet{2011MNRAS.411.1013B}.  The colour scale represents the [Fe/H] value determined by \citet{2011MNRAS.411.1013B} for each of these stars.  They find nine stars inside a radius of 13.2\arcmin from Sextan's centre with similar kinematics to each other, and a metallicity within a scatter of 0.15 dex from [Fe/H] = $-2.6$. Pointing out that the average error in metallicity on these stars is 0.29 dex, they suggest that these stars once belonged to a single stellar population.

\begin{figure}
\centering
\includegraphics[width=\columnwidth]{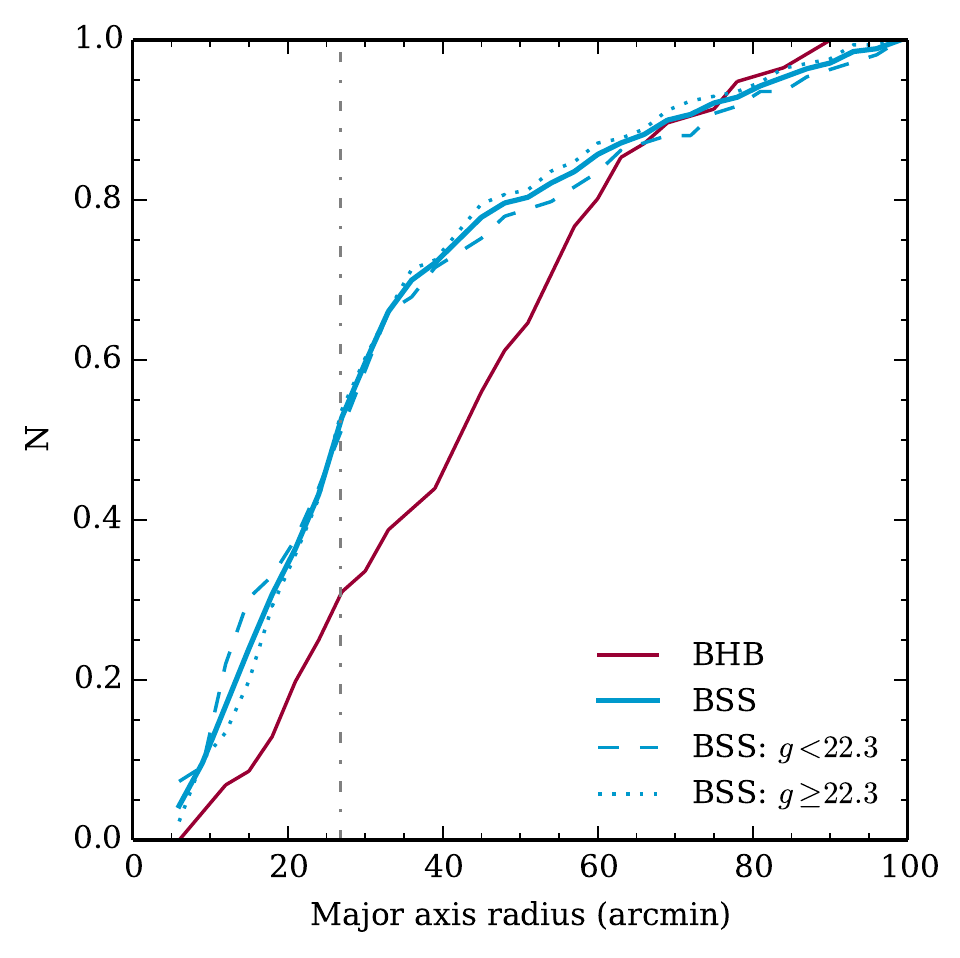}
\caption{Cumulative distribution functions for the BHB and BSS populations (defined in Figure \ref{fig:mask}). The BSS population is also split into  bright and faint groups at $g=22.3$. The grey dash-dot line represents the core radius.}
\label{fig:cdf} 
\end{figure}

Given the differences noted in the distribution of the various stellar populations within Sextans \citep{2003AJ....126.2840L}, we have also performed a brief investigation into the distribution of candidate BSS and BHB stars in our field.  We investigated the cumulative distribution function (CDF) of the radial distribution of the two groups (Figure \ref{fig:cdf}). First, we estimate $\theta$, $\epsilon$ for the two groups, as well as their centroid. For the BHB candidates we find $\theta = 72^\circ\pm15^\circ$ and $\epsilon = 0.20\pm0.06$, and for the BSS candidates $\theta = 40.9^\circ\pm5.4^\circ$ and $\epsilon = 0.36\pm0.02$. The centroids for each group are $-0.086\pm0.009, 0.045\pm0.006$ and $0.02\pm0.01, -0.004\pm0.002$ respectively, measured in degrees from the centre of Sextans. The high uncertainty associated with $\theta$ for the BHB candidates is most likely due to the low number density. However within combined uncertainties, the BSS and BHB populations are largely consistent with the full Sextans sample. Concentric elliptical annuli were placed at evenly spaced intervals of 3\arcmin about each group, and the number of stars inside each annulus counted.  This was then used to create the CDF for each population of stars. A Kolmogorov-Smirnoff  (K-S) test was performed to test the likelihood of the two groups being drawn from the same distribution.  With a $p$-value of 0.38, it is possible the two groups come from separate distributions.

We also looked specifically at the BSS population. \citet{2003AJ....126.2840L} noticed that brighter ($V < 22.3$) stars appeared more centrally concentrated.  We split our BSS population into two at $g=22.3$, plotting the CDF and performing a K-S test. With a $p$-value of 0.998, it is highly likely that the two groups belong to the same distribution.

\section{Discussion and Summary}
\label{sec:disc}

The observation of kinematically cold structures in the centre of Sextans \citep{2004MNRAS.354L..66K,2006ApJ...642L..41W,2011MNRAS.411.1013B}, as well as the suggestion of a dissolved star cluster at its core \citep{2012ApJ...759..111K}, has led to some discussion on whether or not this galaxy is undergoing tidal disruption.  Our analysis of the field surrounding Sextans has revealed significant extended halo-like substructure, extending to a distance of up to $82\arcmin$ (2kpc). Much of this structure appears to be highly statistically significant, with over-densities having $\zeta$ values as high as 26.94, and clear signs of Sextans's stellar population in the colour-magnitude diagram.  We note however, that unlike Hercules \citep{Roderick:2015jj}, the substructure surrounding Sextans appears to be both aligned with, and perpendicular to the major axis.  Figure \ref{fig:summary} shows a summary of our results.  The distribution of over-densities is shown,  colour-coded according to significance.  Different line-widths represent the different detection thresholds, with the fill colour corresponding to the $\zeta$ value at that threshold.  The direction of proper motion of Sextans \citep{2008ApJ...688L..75W} is shown by a black arrow.  

We have shown the candidate BSS population in Figure \ref{fig:summary} with size correlated to brightness; brighter stars appear as larger points.  Previous work has shown that the brighter part of this population is more centrally concentrated than the faint part inside a radius of 22.5\arcmin \citep{2003AJ....126.2840L}.  Our analysis shows that both the bright and faint components of this population are likely to be drawn from the same distribution at the 99.8\% confidence level. We also find that the BSS population is more centrally concentrated than the BHB population, in agreement with \citet{2003AJ....126.2840L}. This fits the idea of \citet{2009ApJ...703..692L} that the star formation history of Sextans varies slightly according to location within the galaxy.

A brief analysis of the central region of Sextans has revealed an over-dense region approximately 9\arcmin to the North-East of the galaxy's centre. Interestingly, this is close to a kinematically cold region detected by \citet{2006ApJ...642L..41W}. Sextans is not the first MW dwarf to show such a feature. The Ursa Minor dwarf shows two apparent `clumps' of stars \citep{1985AJ.....90.2221O,1995MNRAS.277.1354I}, of which the one furthest from centre has been associated with a kinematically cold region \citep{2003ApJ...588L..21K}. It has been suggested that this is a primordial artefact rather than some form of transient substructure, and incompatible with a cusped dark matter halo \citep{2003ApJ...588L..21K}. This has interesting repercussions for $\Lambda$CDM; the appearance of such a feature in Sextans provides an excellent opportunity for further investigation.
 
The radial profile of Sextans is well described by a King model, and shows only tentative evidence of a break radius indicating extra tidal stars.  This is similar to the case of Draco \citep{2001AJ....122.2538O,2007MNRAS.375..831S}, however it is possible that extra tidal features may become apparent further outside the tidal radius; see for example Carina \citep{2000AJ....120.2550M,2005AJ....130.2677M}, although these features are not so prominent in more recent work \citep{2014MNRAS.444.3139M}. Sculptor is another dwarf spheroidal which may \citep{2006AJ....131..375W}, or may not  \citep{2005AJ....130.1065C} show an extra tidal break radius. The large tidal radius observed in Sextans also suggests that the extended stellar halo-structure is bound by the galaxy's gravitational field. Although part of this structure is in proximity to the tidal radius, the majority of structure is well contained. This is consistent with the results of \citet{2009ApJ...698..222P}, where it is noted that there is no evidence to suggest extra tidal features in the radial profile, however they caution that the profile of Sextans was only probed to four times the core radius (approximately 51\arcmin based on their Table 2). According to \citet{2008ApJ...688L..75W}, Sextans is receding from its perigalactic distance of $66_{-61}^{+17}$kpc towards an apogalactic distance of $129_{-33}^{+113}$ kpc.  The large uncertainties make it difficult to discern how eccentric the orbit of Sextans is, and \citet{2008ApJ...688L..75W} state that any orbital eccentricity in the range $0.25 - 0.89$ is within the $95\%$ confidence interval. A more circular orbit would help explain the appearance of the substructure we see surrounding Sextans, as a more eccentric orbit would pass closer \citep[within 5kpc,][]{2008ApJ...688L..75W}  to the Galactic centre and cause more severe tidal disruption.  It has also been shown that a tidally disrupting dwarf may return to a more spherical shape after a few Gyr, depending on the tidal forces involved \citep{2012ApJ...751...61L}. Furthermore, simulations have demonstrated that a galaxy travelling between perigalacticon and apogalacticon will have its tidal tails dissolve, and new ones develop on a relatively short time scale \citep{2009MNRAS.400.2162K}.  These scenarios suggest it is possible Sextans is experiencing tidal disruption, but is presently in a period of morphological transition. The alignment along and perpendicular to the major axis of Sextans over-densities may be an indication the dwarf is in the process of dissolving its tidal tails and developing new ones. 

Given the extended substructure features we have found in this galaxy, as well as variations in the distribution of BHB and BSS populations, it appears that Sextans may have a varied star formation history but is not necessarily undergoing strong tidal disruption.  Although it is possible Sextans is in the midst of a morphological transition due to tidal stirring, it is also possible this galaxy has a circular orbit, exposing it to relatively little influence from the MW. The stars we have identified as belonging to over-densities associated with the Sextans dwarf spheroidal galaxy provide good candidates for spectroscopic follow up.  Velocity measurements for these stars will provide a better picture of the kinematics in the outskirts of this galaxy, and provide us with a better understanding of the nature of its extended substructure.

\section*{Acknowledgements}

The authors wish to thank the referee for thoughtful discussion on the original manuscript. TAR acknowledges financial support through an Australian Postgraduate Award. HJ and GDC acknowledge the support of the Australian Research Council through Discovery Projects DP120100475 and DP150100862. ADM acknowledges the support of the Australian Research Council through Discovery Projects DP1093431, DP120101237, and DP150103294. This project used data obtained with the Dark Energy Camera (DECam), which was constructed by the Dark Energy Survey (DES) collaborating institutions: Argonne National Lab, University of California Santa Cruz, University of Cambridge, Centro de Investigaciones Energeticas, Medioambientales y Tecnologicas-Madrid, University of Chicago, University College London, DES-Brazil consortium, University of Edinburgh, ETH-Zurich, Fermi National Accelerator Laboratory, University of Illinois at Urbana-Champaign, Institut de Ciencies de l'Espai, Institut de Fisica d'Altes Energies, Lawrence Berkeley National Lab, Ludwig-Maximilians Universitat, University of Michigan, National Optical Astronomy Observatory, University of Nottingham, Ohio State University, University of Pennsylvania, University of Portsmouth, SLAC National Lab, Stanford University, University of Sussex, and Texas A\&M University. Funding for DES, including DECam, has been provided by the U.S. Department of Energy, National Science Foundation, Ministry of Education and Science (Spain), Science and Technology Facilities Council (UK), Higher Education Funding Council (England), National Center for Supercomputing Applications, Kavli Institute for Cosmological Physics, Financiadora de Estudos e Projetos, Funda\c{c}\~{a}o Carlos Chagas Filho de Amparo a Pesquisa, Conselho Nacional de Desenvolvimento Cient'fico e Tecnol—gico and the MinistŽrio da Cincia e Tecnologia (Brazil), the German Research Foundation-sponsored cluster of excellence ``Origin and Structure of the Universe" and the DES collaborating institutions.

%%%%%%%%%%%%%%%%%%%%%%%%%%%%%%%%%%%%%%%%%%%%%%%%%%

%%%%%%%%%%%%%%%%%%%% REFERENCES %%%%%%%%%%%%%%%%%%

% The best way to enter references is to use BibTeX:

\bibliographystyle{mnras}
\bibliography{Sextans} % if your bibtex file is called example.bib

% Don't change these lines
\bsp	% typesetting comment
\label{lastpage}
\end{document}